\documentclass[12pt,a4]{article}
\usepackage{amsmath}
\newcommand{\eq}{\begin{equation}}
\newcommand{\eqe}{\end{equation}}
\newcommand{\Ma}{\mathcal{M}}
\newcommand{\Te}{\mathcal{T}}
\newcommand{\Hi}{\mathcal{H}}
\newcommand{\Fo}{\mathcal{F}}
\newcommand{\pa}{\partial}
\newcommand{\ga}{\gamma}
\newcommand{\vt}{\vartheta}
\newcommand{\Th}{\Theta}
\newcommand{\la}{\lambda}
\newcommand{\La}{\Lambda}
\newcommand{\om}{\omega}
\newcommand{\de}{\delta}
\newcommand{\si}{\sigma}
\newcommand{\al}{\alpha}
\newcommand{\ep}{\epsilon}

\newcounter{saveeqn}

\begin{document}
\title{Dirac Quantum Field on Curved Spacetime:\\
        Wick Rotation }
 
\author{Volkhard F. M{\"u}ller\footnote{\ vfm@physik.uni-kl.de}\\
Fachbereich Physik, Technische Universit{\"a}t Kaiserslautern\\
D-67653 Kaiserslautern, Germany }  

\date{ }

\maketitle

\begin{abstract}
 For a linear Dirac field on a globally hyperbolic static
  space-time the analytic continuation of its Wightman functions
  (Green functions) to Schwinger functions and back at zero and
  finite temperature is shown.
 \end{abstract}

\newpage

\section{Introduction }
The Euclidean formulation of relativistic quantum field theory allows
to study concrete models by an approach based on functional integration,
which has lead to a wide variety of both rigorous and perturbative
constructions. The connection between the Euclidean Schwinger functions
and the Wightman distributions (Green functions) on Minkowski space has
been established long ago by Osterwalder and Schrader [OS2], making use, 
among other properties, of the Poincar\'e symmetry present in the
relativistic theory. Considering quantum fields on a Lorentzian space-time
manifold, this symmetry is lacking, which prevents a direct adaption
of the relations valid on Minkowski space. Restricting to space-time 
manifolds, the metric of which is invariant under time translations,
however, the analytic continuation in the time variable can be performed. \\
The purpose of this paper is to show for a linear Dirac field on a
general static space-time the analytic continuation of its Wightman
distributions (Green functions) by a `` Wick rotation'' to Schwinger 
functions and back. Furthermore, proceeding analogously, the thermal
equilibrium functions are considered, first introducing the real-time
distributions by definition and identifying them by the KMS condition
shown to hold. Aiming only at a \emph{Euclidean formulation} of a
(Lorentzian) quantum field theory, no attempt is made to construct a
Euclidean Dirac field operator acting in a related Euclidean Fock space
and yielding the Schwinger functions in the form of vacuum expectation
values, as has been achieved on Minkowski space,[OS1]. Recently Jaffe
and Ritter [JR], considering a ``static'' Riemannian manifold, have 
demonstrated that the inverse of the massive Dirac operator is
reflection positive, thus providing the framework for a quantisation of
the Dirac field. Moreover, within the algebraic approach to quantum
field theory, various properties of the field algebra of a linear
Dirac field on non-static Lorentzian manifolds have been
worked out,[Di],[Ho],[Kr].

The paper is organized as follows: In Section 2 we collect elements of the
classical Dirac equation on a Lorentzian manifold, mainly following [Di],
which are used in the sequel. Aiming at a quantum theory, we restrict 
in Section 3 to a general static manifold. Bringing the Dirac equation 
into the form of a Schr\"odinger equation, we obtain a potential Hamilton
 operator, shown to be a symmetric operator in a Hilbert space 
 related to the metric. We assume that it has a unique self-adjoint
 extension. Cook's method of second quantization [Co] is used in 
 Section 4 to convert the deficient quantum theory into a consistent      
 quantum field theory of the Dirac field, which acts in
 a particle-antiparticle Fock space,
 by attributing the Hamilton operator obtained before,
  which is unbounded from below, to a particle-antiparticle pair. 
 In Section 5 we start from the 2-point Wightman functions, which
 determine the linear theory considered, and analytically continue
 their spectral representation to complex values of the time variables.
 At imaginary time a positivity property is exhibited. Moreover,
 proceeding similarly as Fulling and Ruijsenaars [FR] in the case of
 a scalar field, we deduce a single holomorphic function of complex
 time, which provides the initial Wightman functions as boundary
 values and at purely imaginary time the related Schwinger functions.
 In Section 6 thermal 2-point Wightman functions are defined and 
 analytically continued to complex time with analogous steps as in the
 preceding section. The KMS condition is shown to hold. These
 analytically continued functions again form a single holomorphic
 function, which in this case is antiperiodic in imaginary time and
 provides at purely imaginary time the Schwinger functions. 
\section {The Dirac equation on curved spacetime}
Although in the quantum domain we aim at a static Lorentzian
manifold, in these preparatory classical steps we do not yet
restrict to this particular class. 
i) Let $\Ma$ be a differentiable manifold, connected and 
oriented, $ \rm{dim}\, \Ma = 4 $,
with chart $ (\mathcal{U}, \varphi) $ and local coordinates $ x $ ,
\begin{equation} \label{a1}
 \varphi : \,\mathcal{U} \subset \Ma \longrightarrow \mathbf{R}^4 
        \ni x \equiv \{x^{\mu}\} =(x^0, x^1, x^2, x^3)\, .
 \end{equation} 
 providing in the space of vector fields $ \,\Te^1_{\,\,0}(\Ma) $
 and covector fields (1-forms) $ \Te^{\,0}_{\,\,1}(\Ma) $ 
 the natural dual basis systems
 \begin{equation} \label{a2}
 \{\, \pa_{\mu} \equiv \frac{\pa}{\pa x^{\mu}}\,\}_{\mu =
 0,\dots, 3} \subset \Te^1_{\,\,0} \, ,
 \quad \{ dx^{\mu}\,\}_{\mu = 0,\dots, 3} \subset \Te^{\,0}_{\,\,1} \, : \quad
 \big \langle dx^{\mu}, \pa_{\nu}\big \rangle \, = \, \delta^{\,\mu}_{\,\nu}\,.
 \end{equation} 
 ii) There is a Lorentzian metric
 \begin{equation} \label{a3}
 g = g_{\mu \nu}(x)\, dx^{\mu}\otimes dx^{\nu} \,
 \subset \Te^{\, 0}_{\,\,2} (\Ma) 
 \end{equation}
 with a symmetric nondegenerate 2-cotensor $ g_{\mu \nu}(x)\,$.
 The summation convention on a pair of equal upper and
 lower indices is used, as always 
  in the sequel. Due to the properties of the matrix $ g_{\mu \nu}(x)\,$ there
  exists $ ( c^{\mu}_a (x) ) \in GL(4,R) $ such that 
  \begin{equation} \label{a4}
   c^{\,\mu}_{\,a}(x)\, g_{\mu \nu}(x)\, c^{\,\nu}_{\,b}(x) = \eta_{\, a b}\, ,
   \end{equation}
 where $\,  (\eta_{\, a b}) = \rm{diag}\,(+1, -1, -1, -1) \,$ is the
  Minkowski metric. For the frame indices $ a,b, \cdots \in \{0,\cdots,3\} $
  chosen from the beginning of the Roman alphabet 
  we also use the corresponding summation convention. Defining
 \begin{equation} \label{a5}
 \begin{split}
 g^{\mu \la}(x) \,g_{\la \nu} (x) := \de^{\mu}_{\nu},
 \quad \eta^{\,a b} \eta_{\,b c} := \de^{a}_c \,, \\
 \quad {\vt}^{\,a}_{\mu}(x)\, c^{\nu}_a (x) := \de^{\nu}_{\mu} 
 \,, \, \rightarrow \,\,\vt^{\,a}_{\mu}(x)\, c^{\mu}_b (x) = \de^a_b \,,
 \end{split}
 \end{equation}
 we can invert (\ref{a4})
 \begin{equation} \label{a6}
 g_{\mu \nu}(x) = \eta_{\,a b} \, \vt^{\,a}_{\mu}(x)\,\vt^{\,b}_{\nu}(x)\,,\,\,
 \rightarrow \, g^{\,\mu \nu}(x) =
 \eta^{ a b} c^{\,\mu}_{\,a}(x) c^{\,\nu}_{\,b}(x)\, ,
 \end{equation}
 and introduce in place of (\ref{a2}) new basis systems of vector fields
 and 1-forms, respectively, $a = 0, . .\,, 3 $,
 \begin{equation} \label{a7}
  e_a = c^{\mu}_a(x)\, \pa_{\mu} \,,\quad 
     \Th^{\,a} = \vt^{\,a}_{\mu}(x) \, dx^\mu \,, 
 \end{equation}  
 implying
 \begin{equation}\label{a8}
 \Th^a ( e_b ) \equiv \langle \Th^a , e_b \rangle = \de^a_b \,, \qquad
  g = \eta_{\,a b} \, \Th^a \otimes \Th^b \, ,
  \end{equation}   
  hence providing an orthonormal basis in $\Te^1_{\,\,0} $.   \\
  From the Lorentzian metric (\ref{a3}) follow the associated
  Levi-Civita connection 1-forms
  \begin{equation} \label{b1}
  w^{\la}_{\,\,\mu}(\cdot) \,= 
  \Gamma^{\la}_{\mu \nu}(x)\, dx^{\nu} \, \in \Te^{\, 0}_{\,\,1} \,, \quad
   \la, \mu = 0, \dots, 3 
  \end{equation}  
  with the Christoffel symbols 
  \begin{equation} \label{b2}
   \Gamma^{\la}_{\mu \nu}(x)\,= \, \frac{1}{2}\,\, g^{\la \si}(x)\,
   \{ \pa_{\nu}\, g_{\si \mu}(x) +\pa_{\mu}\, g_{\si \nu}(x) -
 \pa_{\si}\, g_{\mu \nu}(x) \}
    = \Gamma^{\la}_{\nu \mu}(x)\,.
  \end{equation}
  The covariant derivative $ \nabla_{V} $ in direction of a vector field
  $ V = v^{\mu}(x)\, \pa_{\mu} $ is determined when defined on a basis or,
  equivalently, on a cobasis,
   \begin{equation} \label{b3}
    \nabla_{V} \, \pa_{\mu}\, = \, w^{\la}_{\,\,\mu}(V)\, \pa_{\la} \,,\,\, \quad 
    \nabla_{V} \, dx^{\la}\, =\, -\, w^{\la}_{\,\,\mu}(V)\, dx^{\mu} \,. 
  \end{equation}
  Herefrom follows the covariant derivative of the orthonormal basis (\ref{a7}) as
  \begin{equation} \label{b4}
   \nabla_{V}\, e_{\,b} \,= \, \om^a_{\,\, \,b} (V) \, e_{\,a} \, , \, \quad 
   \nabla_{V}\, \Th^a \,= \,-\, \om^a_{\, \,\,b} (V) \, \Th^b \, ,
  \end{equation}
   with transformed connection forms
   \begin{equation} \label{b5}
    \om^a_{\,\,\,b} (V) \, =
 \, \vt^{\, a}_{\la}\, w^{\la}_{\,\,\mu}(V)\, c^{\mu}_{\,b} 
                 + \vt^{\, a}_{\la}\,( V c^{\la}_{\,b}) \,.
  \end{equation}
    Using successively (\ref{b5}), (\ref{b1}), (\ref{a7}) provides
   the relation
 \begin{equation} \label{b6}
      \om^{\,a}_{\,\,\,b}( e_{\,a})\, =  \, c^{\,\la}_{\,b\,;\,\la}\,
                               = \,{\rm div} \, e_{b} \,,
  \end{equation}
  the semicolon denoting the covariant derivative.
  
  A lucid presentation of the classical Dirac equation on a manifold
  is contained in [Di]. Here we collect some elements necessary for a
  quantum version, thereby setting the notation used.\\
  The generators of the Dirac algebra on Minkowski space (Dirac matrices)
   $ \ga^a \in Mat_4(C)\, , a = 0, 1, 2, 3 ,$  satisfy 
   \begin{equation} \label{d1}
    \ga^a\, \ga^b \,+\, \ga^b \, \ga^a \, = \, 2\, \eta^{\, a b}\,\rm{id}\, .
    \end{equation}  
  In addition, we require the property on Hermitean conjugation 
  \begin{equation} \label{d2}
   ( \ga^a )^* \, = \, \ga^0 \,\ga^a \,\ga^0 \,, \quad a=0, \cdots, 3 \,,
  \end{equation} 
  but impose no further constraint, unless stated explicitly. \\
  There is a two-to-one homomorphism of the Lie group $\,Spin(1,3)\ni S \,$
  onto the full Lorentz group $\, O(1,3) \ni \Lambda\,$,
  \eq \label{d3}
   \det S = 1, \qquad S^{-1} \ga^a S \, = \, \Lambda^a_{\,\,\, b}\, \ga^b \,.
  \eqe  
 In the sequel we only consider the restriction of (\ref{d3}) to the respective
 subgroups connected to the identity, i.e.  $\,S \in Spin_0(1,3)\,$ and 
 $\,\Lambda \in SO(1,3)^{\uparrow} \,$. ($ Spin_0(1,3) $ is 
  isomorphic to $\, SL(2,C)$.) \\
 Given any differentiable function 
  $\,\La : \Ma \rightarrow  SO(1,3)^{\uparrow}\,$,
 a local 'gauge' transformation of the respective basis systems 
  (\ref{a7}),(\ref{a8}) in $ \Te^{\,0}_{\,\,1}(\Ma) $ and $\Te^1_{\,\,0}(\Ma) $, 
    \begin{equation} \label{d4}
    (\Th\,')^a : = \La^a_{\,\,b} (x) \, {\Th}^b  \,, \qquad
    e '_{\,b} : =  e_{\,c} ( \La^{-1} (x))^c_{\,\,b} 
    \end{equation}
 yields equivalent basis systems. \\
 A Dirac spinor field on a manifold is a function $ \psi :
 \Ma \rightarrow C^4 \,$,
 which is required to transform as a Dirac spinor with respect 
 to a local gauge transformation
 of the frames, (\ref{d4}). Thus, there is assumed an associated function
 $\, S : \Ma \rightarrow Spin_0(1,3) \,$ such that the homomorphism (\ref{d3}),
 $ S\rightarrow \La\,$,
 holds locally. Then, introducing a moving frame $\,E_A(x), A=1,..,4, $ 
 in the space of spinor fields, $ \psi(x) = \psi^A (x) E_A(x) \,$ (summation
 convention), the companion transformation to (\ref{d4}) is given by
 \eq \label{d5}
 (\psi\,')^A = S(x)^A_{\,\,B}\,\psi^B \, , \qquad
   E\,'_A = E_C (S(x)^{-1})^C_{\,\,A}\, .
 \eqe
 Depending on the global topological structure of $\Ma $ , however, there may
 arise an obstruction to ``lifting'' the function $ \, \La\,$ from the
 non-simply connected group $ \,SO(1,3)^{\uparrow} \,$ to its
 simply connected covering group $ Spin_0(1,3) $. These problems,
 dealt with in [Is], are not pursued here. 
 \footnote{ On a globally hyperbolic Lorentzian manifold this lifting
  can be shown to exist, [Ge].} \\
 The components of the Dirac-adjoint spinor are defined as
 \eq  \label{d6}
  \psi^{+}_A := \overline{(\ga^0)^A_{\,\,B} \psi^B }
    = \sum_{B} \overline{\psi^B} (\ga^0)^B_{\,\, A} \,.
 \eqe  
 Under a frame transformation (\ref{d5}) they transform
 as $\, (\psi ')^{+}_A = \psi^{+}_C\, (S^{-1})^C_{\,\, A}\,$,
 i.e. as the components of a cospinor, because of
 $\, \ga^0 S^{\,*} \ga^0 = S^{-1} \,$ implied by (\ref{d3}),(\ref{d2}).  
 In the space of cospinors a dual moving frame $ E^A, A=1,..,4,$
 can be introduced, $ E^A (E_B) = \delta^A_B \,$,
 then $\, \psi^{+} = \psi^{+}_A E^A $. \\
 \noindent
 From the connection forms (\ref{b4}) of the orthonormal frames
 via the map (\ref{d3}) follows the spin connection form, with
 $\, \om_{ a c}(V):= \eta_{ a b}\, \om^b_{\,\, c}(V) ,$  
 \begin{equation} \label{d7}
   W(V)\,=\, \frac{1}{2} \, \om_{a c} (V) \, \si^{\, a c}\, ,\qquad
   \si^{a b} \, = \, \frac{1}{4}\, ( \ga^a \, \ga^b - \ga^b\, \ga^a ) \,,
 \end{equation} 
 determining the covariant derivatives of the spinor and
 cospinor frames, respectively, in direction of the 
 vector field $ V$, 
 \eq \label{d8}
  \nabla_{V} E_A = E_B \, W(V)^B_{\,\,A} \,, \qquad
  \nabla_{V} E^A = - W(V)^A_{\,\,B}\, E^B \,.
 \eqe 
 Denoting by $\, \nabla_a\,$ the covariant derivative in
 the direction $\,V = e_a\,$, (\ref{a7}), the Dirac equation
 for a spinor field $\, \psi = \psi^A\, E_A \,$ reads
 \eq \label{d9}
   ( - i \ga^a \nabla_a + m )\,\psi \, = 0 \,.
 \eqe 
  Therefrom follow via the Leibniz rule and (\ref{d8}) the
  equations to be fulfilled by the spinor components,
 \begin{equation} \label{d10}
   \big( - i \ga^a \big( e_a \, +\, W(e_a)\big) + m \big)^A_{\,\,B} 
          \, \psi^B \, = \, 0\,.
 \end{equation} 
 Equivalent with these equations the components of the
 Dirac-adjoint spinor satisfy
 \eq \label{d11}
  0\,=\, \psi^{+}_A \big(\, i (\,\stackrel{\leftarrow}{e_a}- W(e_a))\,\ga^a
             + m \big)^A_{\,\, B}\,\,.  
  \eqe 
 Given two solutions $\,(\psi^A), (\chi^A)\,$ of the Dirac equation (\ref{d10}),
 or correspondingly of (\ref{d11}), the bilinear current  
 \eq \label{d12}
 J\,= \, j^{\,\mu}\,\partial_{\mu} \,, \qquad
    j^{\,\mu}:= \, c^{\mu}_a \, \psi^{+}_A (\ga^a)^A_{\,\, B}\, \chi^B
 \eqe
 is covariantly conserved,( observe (\ref{b6})),
 \eq \label{d13}
    { \rm div} J \, = \,j^{\,\mu}_{\,;\, \mu}\, = \, 0 \, .
 \eqe      
 Furthermore, the Dirac operator $\, \ga^a \nabla_a \,$ satisfies the
 Lichnerowicz identity [Li],  
 \eq \label{d14}
 ( - i \ga^a \nabla_a + m )\,(  i \ga^b \nabla_b + m )\,\psi \, = 
  ( \eta^{a b} \nabla_a \nabla_b -  \frac{1}{4}\, R + m^2 )\, \psi \,,
 \eqe
 where $\,R\,$ denotes the scalar curvature of the manifold. The
 principal part of the spinor wave operator
 $\, \eta^{a b} \nabla_a \nabla_b \,$ 
 appearing on the r.h.s. is diagonal and
 coincides with the principal part of the scalar Klein-Gordon operator.   
  \section{Quantum Theory}
 In the sequel we restrict to a \emph{static} Lorentzian manifold
 $\, (\Ma, g) $, hence, topologically $\, \Ma = R \times \Sigma \,$.
 With local coordinates $ x = ( t, \vec x) $ the metric (\ref{a3})
 has components of the form
 \eq \label{q1}
 g_{0\,0} = ( q(\vec x) )^{-2} , \quad \mu, \nu \in \{1,2,3\}:
   \,\, g_{0 \nu} = 0 , \quad g_{\mu \nu} = - h_{\mu \nu}(\vec x)\,,
 \eqe  
 where $\, q(\vec x) > 0\, $ and $ \,(h_{\mu \nu}) \in Mat_3(R)\, $ 
 is positive definite. 
 This metric shows the formal reflection symmetry $ t \rightarrow - t $
 and $ \partial_t $ is a Killing vector field. Upon requiring in addition\\
 A1) $ \quad 0 < c_1 < q(x) < c_2 < \infty \, ,$ \\
 A2) $ \quad (\Sigma, h )\,$ is a complete Riemannian (d=3) manifold,\\
 the space-time manifold $\,(\Ma, g)$, (\ref{q1}), is globally 
 hyperbolic, [Ka]. For simplicity $\,(\Ma, g)$ is taken to be 
 $ \mathcal{C}^{\infty} $. The orthonormal basis
 vector fields (\ref{a7}) emerging from (\ref{q1}) are
 \eq \label{q2}
 c^{\mu}_0 = q(\vec x)\, \delta^{\mu}_0 , \qquad r, s \in \{1,2,3\}:
  \,\, c^0_r = 0 , \quad c^{\mu}_r\, h_{\mu \nu}( \vec x)\, c^{\nu}_s
     = \delta_{r s}\,,
 \eqe
 implying the related dual basis according to (\ref{a5}). \\
 We now refine our notation and denote frame indices which are confined to
 the (spatial) values $1,2,3 $ by letters $\,r, s, ... $ from the end of
 the Roman alphabet and also use a summation convention covering
 these values only.
 From (\ref{q1}) follow the Levi-Civita connection forms (\ref{b5}),
 \begin{equation} \label{q3}
 \om^0_{\,\,\,s}(e_0) = - (\partial_{\mu} \ln q )\, c^{\mu}_{\,s} \,,
    \quad \om^r_{\,\,\,s}( e_0) = 0 \,, \quad 
  \om^0_{\,\,\,s}( e_v) = 0 \,, \quad   \om^r_{\,\,\,s}( e_v) \not= 0 \,.
 \end{equation}
 Moreover, observing the Riemannian volume density 
 $\, |g|^{1/2} = q^{-1}\,|h|^{1/2}\,$ on $ \Ma $, we obtain from
 (\ref{b6}) together with $ \om^0_{\,\, r}(e_{ 0}) $ from (\ref{q3}),
 \eq \label{q4}
  \om^s_{\,\,\,r}( e_s) = {\rm div}_3 \,c_r := 
        |h|^{-1/2}\, \partial_{\la} ( |h|^{1/2} c^{\la}_{\,r})\,.
 \eqe  
 The spin connection forms (\ref{d7}) follow as 
 \eq \label{q5}
    W(e_0) = - (\partial_{\mu} \ln q )\, c^{\mu}_{\,s}\,\si^{ 0 s}\, ,\quad
   W( e_v) = \frac{1}{2} \, \om_{ r s} (e_v) \,\si^{ r s} \,.
 \eqe 
 We write the Dirac equation (\ref{d10}) as an evolution equation 
 to be read as a matrix
 equation, the spinor components $\,(\psi^B)\,$ forming a column matrix,
 and introduce the time-honoured Hermitian matrices
  $\, \alpha^r = \ga^0 \ga^r , \, \beta = \ga^0$, 
 \eq \label{q6}
 \begin{split}
   i q(\vec x)\, \partial_t \psi \, =
    \,( K - i W(e_0))\,\psi \,,\qquad \qquad \,\\
   K: = -i\,\al^r e_r -i\,\al^r W(e_r) + m\beta \,.
   \end{split}
 \eqe   
 A Hilbert space $ \mathcal H $  formed of Dirac spinors $ \psi, \chi, \dots $ 
 emerges from the inner product (at fixed $t$),
 \eq \label{q7}
  (\psi, \chi) : = \int_{\Sigma} d^3 x \,|h|^{1/2} \sum_{A=1}^4 
    \overline{\psi^A}\, \chi^A\,,
 \eqe
 suggested by the conserved current (\ref{d12}), upon completion. 
 We first notice, with 
  $ \,\psi, \chi \in {\mathcal C}^{\infty}_0 (\Sigma)^{\,4} $ , i.e. 
   smooth functions of compact support, 
 \eq \label{q8}
 (\psi, W(e_0)\, \chi) = ( W(e_0)\, \psi,\, \chi)\, , \qquad 
 (\psi, K\,\chi) = ( K \, \psi, \chi) \,,
 \eqe 
 using (\ref{q4}) in the latter case. The multiplication  operator
 $\,q\,$ defined on $\,\Hi\,$ is bounded, positive, self-adjoint and
 invertible. Furthermore, on the domain 
 $ \, {\mathcal C}^{\infty}_0 (\Sigma)^{\,4} \subset \Hi\, $  holds
 the commutation relation
 \eq \label{q9}
 \big[\, K\,, \frac{1}{q}\, \big] \,= \, - 2\, i \,\frac{1}{q}\, W(e_0)\,.
 \eqe 
 Hence, the operator
 \eq \label{q10}
  H\, = \, \frac{1}{q}\, \big(\, K - i W(e_0) \big)
     = \, \frac{1}{2}\, \Big(\,\frac{1}{q}\, K + K \,\frac{1}{q}\Big)
 \eqe
  is symmetric on this domain, too,
  \eq \label{q10a}
  (\psi, H\,\chi) = ( H \, \psi, \chi) \,.
  \eqe
   We \emph{assume}, that it has a unique
  self adjoint extension, called the Hamilton operator in  $ \,\mathcal H \,$
  and denoted again by $\,H\,$ in the sequel. 
 For any initial value $\, \psi_0 \in {\mathcal H}\,$ at $ t=0 $ then
 \eq \label{q11}
   \psi (t, \vec x) = e^{ -i t H}\, \psi_0
 \eqe
 is a solution of (\ref{q6}) in $\,{\mathcal H}\,$, and the inner
 product (\ref{q7}) of two such solutions does not depend on time.
  The operator $ H $, however, is not bounded from below, hence
  does not allow to consider $\, {\mathcal H}\,$ as a one-particle
  space. But $\, {\mathcal H}\,$  has an
  inherent conjugation symmetry. To exhibit this symmetry, one observes 
  that the generators of the Dirac algebra (\ref{d1}) in the standard 
  representation \footnote{
  \begin{equation} 
  \ga^0 =
  \left( \begin{array}{cc}
   \si_0 & 0 \\
   0 & - \si_0
   \end{array} \right ) \, , \quad
  \ga^k = 
  \left( \begin{array}{cc}
   0 & \si_k \\
   - \si_k & 0
   \end{array} \right )\,,
   \quad k = 1,2,3 \, ,
   \end{equation}
   with $ \,\si_0\, $ the $ 2 \times 2 $ unit matrix and $ \,\si_k \,$ 
   the standard Pauli matrices.}
  ( or in the Weyl representation) have  in addition to (\ref{d2}) the property 
 \eq \label{q12}
 a = 0, 1, 3: \quad \overline{\ga^a} = \ga^a \,,\qquad 
           \overline{\ga^2} = - \,\ga^2 \,.
 \eqe  
 Henceforth, we require this property, too, shown by any representation
 resulting from the stated ones by a \emph{real unitary} similarity 
 transformation. Given such a representation, the real matrix 
 $\, C:= i \ga^2 \,$, hence $\, C = C^* = C^{-1}\,$, implies 
 \eq \label{q13}
    C \, \overline{\ga^a}\, C^* = - \, \ga^a \,, \quad a = 0,..,3 \,.
 \eqe
 Defining then in $ \Hi $ the antiunitary involution $ A $,
 \eq \label{q14}
 \psi \in \Hi : \qquad A \,\psi\, = \, C \overline{\psi} \,,
 \eqe
 the Hamilton operator (\ref{q10}) satisfies
 \eq \label{q15}
      A\, H\, = \, - \, H \,A \,.
 \eqe
 The spectral representation of $ \,H\, $ provides in $ \,\Hi\, $ 
 the orthogonal projection operators $ P_{+} > 0 \,$ and
 $ P_{-} < 0 \,$ corresponding to the positive and negative part
 of the spectrum, respectively, $\, P_{+} + P_{-} = id \,$, a separating
 gap is tacitly assumed.
 Defining $ \Hi_{+}= P_{+}\Hi\,\,,\Hi_{-}= P_{-}\Hi\,$, the
 Hilbert space decomposes as a sum of two orthogonal subspaces, 
 \eq \label{q16}
  \Hi \,= \, \Hi_{+} \oplus \,\Hi_{-} \, .
 \eqe 
 Furthermore, the conjugation (\ref{q13}),
 \eq \label{q17}
    A \,: \, \Hi_{+}\, \rightarrow \,\, \Hi_{-}
 \eqe
 is a bijective map. 
 
 \section{ Quantum Field Theory}
 The deficient quantum theory of the previous section is elevated
 to a consistent quantum field theory using Cook's method [Co] of
 second quantization. \footnote{early applications of this method
 are e.g. [Bo], [DM].} In addition to the Hilbert space $ \Hi $,\,(\ref{q16}),
 a ``physical'' Hilbert space $ \,\Hi'\, $ is introduced,
 \eq \label{f1}
       \Hi' \, = \, \Hi_{p} \oplus \Hi_{{\bar p}} \,, 
 \eqe 
 related to $\, \Hi\,$ by the map 
 \eq \label{f2}
 \nu : \, \Hi \, \rightarrow \, \Hi'\,,\qquad
 \nu = I_p \,P_{+} \,+\, I_{{\bar p}}\, A\,P_{-} \,,
 \eqe
 with identification maps $ I_p $ on $ \Hi_p $ and
 $ I_{{\bar p}} $ on $ \Hi_{{\bar p}} $.
 The Fock space is built on $\Hi' $ in the standard way,
 \eq \label{f3}
 \mathcal{F} = \mathbf{C} \Omega \oplus \sum_{n=1}^{\infty} (\Hi')^{\otimes n}
   |_{\rm {as}}
 \eqe
 with the one-dimensional subspace spanned by the 
 vacuum vector $\Omega $ and a direct sum
 of totally antisymmetrized tensor products of $\Hi'$. 
 Let $ \langle \cdot\, , \cdot \rangle $ denote the scalar product 
 of $ \Fo $, (\ref{f3}), and the vacuum vector $\,\Omega\, $ being normalized,
 $\, \langle \Omega, \Omega \rangle = 1 $. 
 In $ \mathcal{F} $
 annihilation operators $\, c(f')\,$ with $\, f ' = \nu f, f \in \Hi $ 
 are defined, their adjoint $ c^{*}(f')$ acting as creation operators.  
 We split in notation, cf.(\ref{f2}), and denote by $ a( P_{+}f ) $
 and $ b( A P_{-}f) $
 the annihilation operator of a particle and an antiparticle, respectively.
 These operators and their adjoints satisfy the anticommutation relations,
 $ f,g \in \Hi ,$
 \begin{eqnarray} \label{f4}
 \{ a(P_{+}f)\, ,\, a^{*}(P_{+}g)\} & = & (\, P_{+}f, g) \\
 \{ b( A P_{-}f)\, ,\, b^{*}( A\, P_{-}g) \}  & = &
    (\, A P_{-}f,\, A P_{-}g ) = \overline{( P_{-}f,\, g)} \, , \nonumber
 \end{eqnarray}       
 all other anticommutators vanishing. Given these operators acting in
 $ \Fo $, the (smeared) field operator at time $ t=0 $ is defined as
 \eq \label{f5}
 \Psi (f):= a( P_{+}f) + b^{*}(A P_{-}f) \,,
 \eqe
 satisfying
 \eq \label{f6}
 \{ \Psi (f)\, , \Psi^{\,*} (g) \} = (f, g) \, .
 \eqe 
 One should notice that in the definition (\ref{f5}) of the 
 smeared field operator $ f \in \Hi $ acts antilinearly on the
 operator-valued distribution $ (\Psi^A(\vec x)) $, 
 \eq \label{f7}
 \Psi (f) = \int d^3 x |h|^{1/2} \sum_{A} 
             \overline {f^A} ( \vec x) \Psi^A (\vec x) \, .
 \eqe  
 Finally the time dependence of the field operator follows from (\ref{f5}) as
 \eq \label{f8}
 \Psi (t, f) := \Psi ( e^{itH} f) \,,
 \eqe
 implying the linear field equation
 \eq \label{f8a}
  \Psi ( ( i \partial_t + H )\, e^{itH} f)\, = \, 0 \,.
 \eqe
 The anticommutation relations of the time dependent fields
 result from (\ref{f6}) as,
 \eq \label{fac}
 \{ \Psi ( t, f)\, , \Psi^{\,*} (t', f') \} = 
             (f, e^{-i ( t-t')H} f') \, ,
 \eqe 
 the other ones vanishing. Moreover, if we restrict $\, f,f' \in \Hi \,$
 to functions $ f, f' \in C^{\infty}_0 (\Sigma)^4 \,$ such that
 the spacetime domains $ ( t, {\rm{supp}}\, f )$ and 
 $ ( t', {\rm{supp}}\, f' )$ are causally disjoint (c.d.), i.e. there
 is no spacetime point in one of these domains, that can be connected
  with any point of the other by a forward or a backward time-like or 
 light-like curve, then 
 \eq \label{facd}
 \{ \Psi ( t, f)\, , \Psi^{\,*} (t', f') \} = \,0 \,, \,\, {\rm{if}}\,\,
    ( t, {\rm{supp}}\, f ) \,\, {\rm{c.d.}} \,( t', {\rm{supp}}\, f' )\,.      
 \eqe 
 This property eventually is a consequence of the finite propagation speed
 of a hyperbolic differential equation. Although the Dirac equation itself 
 is not hyperbolic, the r.h.s. of the Lichnerowicz identity (\ref{d14}) is,
 herefrom the claim can be deduced, see [Di]. 
 
 In $ \Fo $ the operator $ U_c $ of particle-antiparticle conjugation, defined as
 \eq \label{f9}
  U_c \,\Omega = \Omega \, , \qquad U_c\, \Psi( f)\, U^*_c = 
            \epsilon\, \Psi^* ( A f)
 \eqe
 with a (physically irrelevant) phase factor $ \epsilon\,,\, |\epsilon | = 1\,$,
 and the antiunitary operator $ A $ in $ \Hi\,$ from (\ref{q14}) is a 
 unitary operator. From (\ref{f9}) then follows
 the transformation of the Heisenberg field operator
 \eq \label{f10}
   U_c\, \Psi( t, f)\, U^*_c = \,\epsilon\, \Psi^* ( t, A f)
 \eqe
 \section {Green functions and Schwinger functions} 
 The theory
 considered being linear, all $n$-point Wightman functions are
 sums (with signs) of products of $2$-point functions.
 Furthermore, a non-vanishing $n$-point function requires
 $\, n=2l,\, l \in \mathbf{N}, $ and has to be composed of
 $\,l\, $ factors $ \Psi $ and of $\,l\,$ factors $ \Psi^{*} $.
 The non-vanishing $2$-point functions follow as
 \eq \label{gs1}
 \begin{split}
  \langle \Omega , \Psi(t, f) \Psi^{*}( t', f')\, \Omega \rangle
      = ( f, P_{+}  e^{- i(t-t')H} f' ) \,, \\
 \langle \Omega , \Psi^{*}(t', f') \Psi( t, f)\, \Omega \rangle
      = ( f, P_{-}  e^{- i(t-t')H} f ' ) \,,
 \end{split}         
 \eqe 
 and are related according to the symmetry (\ref{f10}),
 \eq \label{gs2}
 \langle \Omega , \Psi^{*}(t', f') \Psi( t, f)\, \Omega \rangle =  
 \langle \Omega , \Psi(t', A f') \Psi^{*}( t, A f)\, \Omega \rangle \,.
 \eqe 
 The spectral representation (\ref{gs1}) of these $2$-point functions 
 can be analytically continued to complex values of the time
 coordinates, $\, t \rightarrow t+i\tau,\, t'\rightarrow t'+i\tau',\\
 \,\tau, \tau' \in \mathbf{R}$. Defining  $\,z = t-t' + i(\tau-\tau')\,$
 with $\, s:= {\rm{Im}}z = (\tau-\tau')\,$ we obtain from (\ref{gs1})
 by analytic continuation $\,t-t' \rightarrow z\,$ the
  corresponding functions
  $\,F^{(+)}(z; f, f') $, holomorphic in $\, {\rm{Im}}\,z < 0\,$, and  
   $\,F^{(-)}(z; f, f')\,$, holomorphic in $\, {\rm{Im}}\, z > 0\,$. 
 The original $2$-point functions, which are continuous functions
  of $\,t-t'$, appear as boundary values on the real axis, 
 \eq \label{gs3}
 \begin{split}
  \langle \Omega , \Psi(t, f) \Psi^{*}( t', f')\, \Omega \rangle \,
      = \,\lim_{\epsilon \searrow 0} F^{(+)}(t-t'- i\epsilon\, ;f, f')\,, \\
 \langle \Omega , \Psi^{*}(t', f') \Psi( t, f)\, \Omega \rangle
      = \,\lim_{\epsilon \searrow 0} F^{(-)}(t-t'+ i\epsilon\, ;f, f')\,.
 \end{split}        
 \eqe 
 Considering these holomorphic functions $\, F^{(\pm)}( z ; f, f')\,$
 in their respective domains
 at `` imaginary time'' $\, z=is, \,$ we notice on the
 functional diagonal $ \, f = f'\,$ the following positivity properties,
 \eq \label{gs3p}
 \begin{split}
 F^{(+)}( is\,; f, f) \geq 0\,,\quad  \,s < 0\,,\\
  F^{(-)}( is\,; f, f) \geq 0\,,\quad  \,s > 0\,.
  \end{split}
 \eqe 
 Since $\,s= \tau - \tau' $ we observe, that these positivity properties
 hold, if the product of fields entering the 2-point function is ordered
 (from left to right) according to \emph{increasing} values of
 \emph{imaginary time}. In particular, this ordering is given by requiring
 the factor on the right(left) to have positive(negative) imaginary time.
 The resulting positivity then implies \emph{reflection positivity}, which forms
 the basis in an Euclidean approach on a `` static'' Riemannian manifold [JR].
 
 Moreover, restricting now $\, f, f' \in \Hi\,$ to functions
  $\, f, f' \in C^{\infty}_0 (\Sigma)^4 $ which have disjoint
 supports, $ {\rm{supp}}\,f \cap {\rm{supp}}\,f' = \emptyset $, then 
 according to (\ref{facd}) there exists an intervall
  $ |t-t'| < d \equiv d( {\rm{supp}}\,f, {\rm{supp}}\,f')\,$, such that 
 for corresponding values of $ t-t' $ holds
 \eq \label{gs4}
       \lim_{\epsilon \searrow 0} F^{(+)}(t-t'- i\epsilon\, ;f, f')\,+
      \,\lim_{\epsilon \searrow 0} F^{(-)}(t-t'+ i\epsilon\, ;f, f')\,= \, 0\,.
 \eqe  
 Hence, due to a theorem of Painlev\'e, [SW, Theor.2.13]
 or [Hi, Theor. 7.7.1], there is 
 a single function $ F(z\,; f, f')\,$, which is holomorphic in the 
 \emph{cut plane} $\, z \in {\mathbf C} \setminus ({\mathbf R} \setminus I )\,$
 and continuous at the boundary, where $ I $ denotes the real intervall
  $ I=(\,{\rm{Im}}\, z = 0,| {\rm{Re}}\, z | < d )$ ,
  such that
 \begin{equation} \label{gs5}
 F(z\,; f, f') = \left\{ \begin{array}{ll}
    +\,F^{(+)}(z\,; f, f')\,,\quad {\rm{Im}}\, z \leq 0 \,,\\ 
   - \,F^{(-)}(z\,; f, f')\,,\quad {\rm{Im}}\, z \geq 0\,. 
 \end{array} \right.
 \end{equation}
 Because of the cut on the real axis for $\, ( z \in {\mathbf R}, |z| > d )\,$ 
 there generically arise different boundary values from above and below.\\
 In addition, we can define a ``Wick-rotated'' version of this function, 
 \eq \label{gs6}
 F_E( \zeta\,; f, f') := F( i\zeta\,; f, f') \,, \quad \zeta \in {\mathbf C}\,,
 \eqe  
 with a domain of holomorphy resulting in an obvious way
 from that of the function $F$.
 For real values of $\, \zeta $, corresponding to ``imaginary time'',
 then emerges the Schwinger function, 
 \eq \label{gs7}
  s \in {\mathbf R}: \quad  F_E( s\,; f, f') = 
  \, \theta(-s)( f, P_{+} \, e^{s H} f' )\,
     -\,\theta(s)\,( f, P_{-} \, e^{s H} f' ) \,.
 \eqe 
 The Schwinger function (\ref{gs7}) satisfies 
 \eq \label{gs8}
 \partial_s \,F_E( s\, ; f, f') -  F_E( s\, ; Hf, f') 
  \,= \,- \, \delta (s)\, ( f, f') \, ,
  \eqe 
  where the r.h.s., however, vanishes because of the assumed support
  properties of the functions $ f, f' $. We notice, that the function
   (\ref{gs7}) considered autonomously, is also well defined for
    functions $ f, f' \in C^{\infty}_0(\Sigma)^4 $ without the 
    restriction on their supports.\\ 
  From the Schwinger function (\ref{gs7}) 
 one recovers the Wightman functions (\ref{gs3}) by analytic
  continuation, with $ \epsilon > 0 $, 
  \eq \label{gs9}
  \begin{split}
   \lim_{\epsilon \searrow 0} F_E( -\, \epsilon - i(t-t') ;f, f')\,=
   \,\lim_{\epsilon \searrow 0} F^{(+)}(t-t' - i\epsilon\, ;f, f')\,, \\
  \lim_{\epsilon \searrow 0} F_E( \epsilon - i(t-t') ;f, f')\,=
   \,- \,\lim_{\epsilon \searrow 0} F^{(-)}(t-t' + i\epsilon\, ;f, f')\,.
  \end{split}   
 \eqe 
 Wightman functions involving in place of the adjoint field operator
 $\,\Psi^{*}(t, f) \,$ the Dirac-adjoint field
 field operator $\,\Psi^{+}(t, f)\,$ 
 are easily obtained from the former ones. By definition, cp.(\ref{d6}), 
 the Dirac-adjoint of the field operator (\ref{f8}) is given by
 \eq \label{gs10}
 \Psi^{+}(t, f)\, = \,\Psi^{*}(t,\ga^0 f) \,.
 \eqe
 Hence, substituting in (\ref{gs1}) the function $ f' $ by $ \ga^0 f'$ 
 converts the adjoint field operator into the Dirac-adjoint field
 operator according to (\ref{gs10}).\\
 Finally, as a consequence of (\ref{gs9}) the time-ordered product
 \footnote{with standard time ordering of fermion operators} follows,
 \eq \label{gs9a}
 \langle \Omega , T \Psi(t, f) \Psi^{*}( t', f')\, \Omega \rangle =
 \lim_{\epsilon \searrow 0} F_E( - i(t-t')(1-i\epsilon) ;f, f')\,.
 \eqe
 
 The approach includes Minkowski space, of course, as a particular instance. 
 Then we have $\, \Sigma = {\mathbf R}^3 \,$ and the metric (\ref{q1}) has the
 particular form
 \eq \label{gs11}
  q(\vec x) \equiv 1 ,\quad  h_{\mu \nu}(\vec x) = 
    \delta_{ \mu, \nu}\,,  \quad \mu, \nu \in \{1,2,3\}.
 \eqe 
 The Wightman functions (\ref{gs3}), transcribed to involve the 
 Dirac-adjoint field, have the particular form
  \begin{eqnarray} \label{gs12}
  \lim_{\epsilon \searrow 0} F^{(\pm)}
   (t-t' \mp\, i\epsilon\, ;f,\ga^0 f')\,=\qquad \qquad \qquad \\
  \frac{1}{i}\, \int d^3x \int d^3x' \,\sum_{A,B} \overline {f^A(\vec x)}
   \, S^{(\pm)}(t-t', {\vec x}- {\vec x'})^A_{\,\,B} f'^B({\vec x}\,')
  \nonumber
  \end{eqnarray}
  where $\, S^{(\pm)}( x - x')\, $ are the familiar $2$-point functions
   of the  free Dirac field,
   \eq \label{gs13}
   \frac{1}{i}\, S^{(\pm)}(x-x') \,=\, (2\pi)^{-3}\int \, \frac{d^3 p}{2 p^0}
     \,( \ga p \pm m) e^{\mp i p(x-x')}\, ,
   \eqe    
   with $\, p^0 = ( {\vec p} ^{\,2} + m^2 )^{1/2} \,$, see eg. [IZ].  
       
 \section{ Thermal Equilibrium}
 With the notation introduced before, a real parameter $ \beta $, where
 $\, 0 < \beta < \infty\,$, and
 $\, f, f' \in C^{\infty}_0(\Sigma)^4\,$, we define the functions
 \footnote{These definitions are suggested by heuristically manipulating
 the Gibbs formula.}
 \eq \label{th1}
 \begin{split}
  F^{(+)}_{\beta}(z\, ;f, f') := 
 \big( f, P_{+} \,\frac{e^{-izH}}{1+ e^{-\beta H}}f'\big)  
 + \big( f, P_{-}\, \frac{e^{-izH + \beta H}}{1+ e^{\beta H}}f'\big)\,,\\ 
   {\rm{for}}\quad -\beta \leq {\rm{Im}}\, z \leq 0\, ; \\
 F^{(-)}_{\beta}(z\, ;f, f') := 
 \big( f, P_{+} \,\frac{e^{-izH - \beta H}}{1+ e^{-\beta H}}f'\big)  
 + \big( f, P_{-}\, \frac{e^{-izH}}{1+ e^{\beta H}}f'\big)\,,\\ 
 {\rm{for}}\quad 0 \leq {\rm{Im}}\, z \leq \beta.
 \end{split}
 \eqe 
 $ F^{(+)}_{\beta}(z\,; f, f') $ is a holomorphic function in the open strip
 $ -\beta < {\rm{Im}}\, z < 0 $, whereas $  F^{(-)}_{\beta}(z\,; f, f') $ is 
 holomorphic in the open strip $ 0 < {\rm{Im}}\, z < \beta $, and both functions
 are continuous on the respective boundary. \\
  For $\, \beta \nearrow \infty \,$ 
 these functions have as limits the functions $\,F^{(\pm)}\,$ defined
  before after (\ref{gs2}),
 \eq \label{th2}
 \begin{split}
 \lim_{\beta \nearrow \infty} F^{(+)}_{\beta}(z\, ;f, f') = 
  ( f, P_{+} e^{-i z H} f') = F^{(+)}(z ;f, f')\,, \quad {\rm{Im}}\,z \leq 0\,,\\
 \lim_{\beta \nearrow \infty} F^{(-)}_{\beta}(z\, ;f, f') = 
  ( f, P_{-} e^{-i z H} f') = F^{(-)}(z\,;f, f')\,, \quad {\rm{Im}}\,z \geq 0\,.
 \end{split}
 \eqe 
 Furthermore, considering the functions (\ref{th1}) 
  at `` imaginary time'' $\, z=is\,$, we notice on the
 functional diagonal $ \, f = f'\,$, similarly as in (\ref{gs3p}), 
 the positivity properties,
 \eq \label{th2p}
 \begin{split}
 F^{(+)}_{\beta}( is\,; f, f) \geq 0\,,\quad \, \,-\beta \leq s \leq 0\,,\\
  F^{(-)}_{\beta}( is\,; f, f) \geq 0\,,\qquad  \,0 \leq s \leq \beta\,.
  \end{split}
 \eqe 
 Reading again $ \,s=\tau - \tau'\,$ as difference in imaginary time,
 positivity obviously requires an ordering condition to be satisfied:
 $\, 0 < \tau' - \tau < \beta\,$ in the first inequality, and
 $\, 0 < \tau - \tau' < \beta\,$ in the second one.\\
 Moreover, with $\,\beta\,$ fixed, due to their very definition (\ref{th1}) the
 functions $\,F^{(\pm)}_{\beta}\,$ are related in the respective
 closure of the domain of holomorphy,
 \eq \label{th3}
 0 \leq {\rm{Im}}\,z \leq \beta : 
   \quad  F^{(+)}_{\beta}(z - i\beta\,;f, f') = F^{(-)}_{\beta}(z\,;f, f')\,.
 \eqe  
 This is the \emph{Kubo-Martin-Schwinger (KMS) condition} [Ku],[MS] 
 characterising thermal equilibrium
 at temperature $\,T = 1/\beta $, [HHW].\\
  Furthermore, in case of real values of $\,z\,$
  follows from the definitions (\ref{th1}) with $ t-t' \in \mathbf{R} :$
 \eq \label{th4}
  \lim_{\ep \searrow 0}\,F^{(+)}_{\beta}(t-t'- i\ep \,;f, f')
  \,+ \,\lim_{\ep \searrow 0}\,F^{(-)}_{\beta}(t-t' +i\ep \,;f, f')
     \,=\, ( f, e^{-i(t-t') H} f') \,.
  \eqe
 Requiring now the functions $\, f, f' \in C^{\infty}_0 (\Sigma)^4 $ to have
 disjoint supports, $ {\rm{supp}}\,f \cap {\rm{supp}}\,f' = \emptyset $, then 
 as in the previous case (\ref{gs4}) the r.h.s. of (\ref{th4}) vanishes for all
 $\, t-t'\,$ satisfying
 $ |t-t'| < d \equiv d( {\rm{supp}}\,f,{\rm{supp}}\,f')\,$, 
 because of finite propagation speed. ( See the remarks after (\ref{fac}),
 (\ref{facd}).) Hence, similarly as in (\ref{gs5}), the Painlev\'e theorem
 again yields a single function $ F_{\beta}(z\,; f, f')\,$,
  holomorphic in the \emph{cut open strip} 
 $\,( - \beta < {\rm{Im}}\, z < \beta ) \setminus ({\mathbf R} \setminus I )\,$
 with a gate on the real axis given by the intervall 
 $ I=(\, {\rm{Im}}\, z = 0,| {\rm{Re}}\, z | < d\, )$ , 
  \begin{equation} \label{th5}
 F_{\beta}(z\,; f, f') = \left\{ \begin{array}{ll}
    +\,F^{(+)}_{\beta}(z\,; f, f')\,,\quad -
 \beta \leq {\rm{Im}}\, z \leq 0 \,,\\ 
   - \,F^{(-)}_{\beta}(z\,; f, f')\,,\quad \quad
 0 \leq {\rm{Im}}\, z \leq \beta\,. 
 \end{array} \right.
 \end{equation}
 The boundary values on the real axis are continuous, but 
 the cut has to be observed. From the KMS condition 
 (\ref{th3}) then follows that this analytic function $\, F_{\beta}(z\,; f, f')\,$
 satisfies
 \eq \label{th6}
 0 \leq {\rm{Im}}\,z \leq \beta : 
   \quad  F_{\beta}(z - i\beta\,;f, f') = - \, F_{\beta}(z\,;f, f')\,,
 \eqe  
 i.e. is anti-periodic in imaginary time. \\
 The relation (\ref{th6}) can also 
 be used as a definition to extend the function $\, F_{\beta}(z\,; f, f')\,$ 
 by analytic continuation from the strip $ 0 \leq {\rm{Im}}\,z \leq \beta $
 to the adjacent strip $ \beta \leq {\rm{Im}}\,z \leq 2\beta $ and so on,
 as well as proceeding similarly from the strip
  $ -\beta \leq {\rm{Im}}\,z \leq 0 $ 
 downward. Thus results a function  $\, F_{\beta}(z\,; f, f')\,$  
 which is holomorphic in the cut plane 
 $\,z \in {\mathbf C}\setminus L\,$ with the set of cuts
 $\,L = ( t+in\beta\,:\, t \in {\mathbf R}, |t|> d, n \in {\mathbf Z})$, and
 continuous on the boundary, satisfying (\ref{th6}) everywhere.\\

 \vspace{1cm}
\noindent
{\bf References}\\
\begin{itemize}
\item[{[Bo]}] P.J.M. Bongaarts, The electron-positron field coupled to external\\ 
    electromagnetic potentials as an elementary C*-algebra theory,\\
     Ann. Physics {\bf 56} (1970) 108-139
\item[{[Co]}] J.M. Cook, The Mathematics of Second Quantization,\\
         Trans. Amer. Math. Soc. {\bf 74} (1953) 222-245
\item[{[Di]}] J. Dimock, Dirac Quantum Fields on a Manifold,\\
       Trans. Amer. Math. Soc. {\bf 269} (1982) 133-147 
\item[{[DM]}] H.G. Dosch and V.F. M\"uller, Renormalization of Quantum
   Electro- \\dynamics in an Arbitrarily Strong Time Independent External Field,\\
  Fortschr. Phys. {\bf 23} (1975) 661-689       
\item[{[FR]}] S.A. Fulling and S.N.M. Ruijsenaars, Temperature, Periodicity and \\
          Horizons, Phys. Reports {\bf 152} (1987) 135-176  
\item[{[Ge]}] R. Geroch, Spinor structure of space-times in
         general relativity.II,\\
     J. Math. Phys. {\bf 11} (1970) 343-348
\item [{[HHW]}] R. Haag, N. Hugenholtz and M.Winnink,\\ On the Equilibrium
               States in Quantum Statistical Mechanics,\\
                Commun. Math. Phys. {\bf 5} (1967) 215-236    
\item[{[Hi]}] E. Hille, \emph{Analytic Function Theory}, vol.1,\\
              Ginn and Company, New York, 1959 
\item [{[Ho]}] S. Hollands, The Hadamard Condition for Dirac Fields and\\
      Adiabatic States in Robertson-Walker Spacetimes,\\
       Commun. Math. Phys. {\bf 216} (2001) 635-661
\item [{[Is]}] C.J. Isham, Spinor fields in four dimensional space-time,\\
            Proc. Roy. Soc. Lond.{ \bf A.364} (1978) 591-599 
\item[{[IZ]}] C. Itzykson and J.B. Zuber,  \emph{Quantum Field Theory},\\
              McGraw-Hill, 1980
\item [{[JR]}] A.Jaffe and G.Ritter, Reflection Positivity and Monotonicity,\\
               J. Math. Phys.{ \bf 49} (2008) 052301 
\item [{[Ka]}] B.S. Kay, Linear Spin-zero Quantum Fields in\\
          External Gravitational and Scalar Fields,\\
               Commun. Math. Phys. {\bf 62} (1978) 55-70                
\item [{[Kr]}] K. Kratzert, Singularity structure of the two point
                      function of the\\
              free Dirac field on globally hyperbolic spacetime,\\
                Annalen Phys.{ \bf 9} (2000) 475-496                             
\item [{[Ku]}] R. Kubo, 
         J. Phys. Soc. Japan{ \bf 12} (1957) 570-                
\item[{[Li]}]  A. Lichnerowicz, Champs Spinoriels et Propagateurs en \\
   Relativit\'e G\'en\'erale, Bull. Soc. Math. France {\bf 92} (1964) 11-100 
\item [{[MS]}] P.C. Martin and J. Schwinger,
              Phys. Rev.{ \bf 115} (1959) 1342-
\item [{[OS1]}] K. Osterwalder and R. Schrader, Feynman-Kac 
        Formula for Euclidean Fermi and Bose Fields,
              Phys. Rev. Lett. {\bf 29} (1972) 1423-1425               
\item [{[OS2]}] K. Osterwalder and R. Schrader, Axioms for Euclidean Green's \\
               Functions,
              Commun. Math. Phys.{ \bf 42} (1975) 281-305                       
\item[{[SW]}] R.F. Streater, A.S. Wightman,  \emph{PCT, Spin and Statistics,
        and all That}, Benjamin, New York, 1964
 \end{itemize}  

 \end{document}